\documentclass[aps,twocolumn,tightenlines,nofootinbib,superscriptaddress]{revtex4-2}
\usepackage[utf8]{inputenc}
\usepackage{graphicx}
\usepackage{subcaption}
\usepackage{mwe}
\usepackage{mathrsfs}
\usepackage{amsfonts}
\usepackage{setspace}
\usepackage{cellspace}
\usepackage{amsmath,amssymb,bm}
\usepackage[colorlinks=true,linkcolor=blue]{hyperref}
\usepackage{xcolor}
\usepackage{epsfig}
\usepackage{slashed}
\usepackage{caption}
\usepackage{hhline,multirow,tabularx}  %
\usepackage{dcolumn}    %
\usepackage{url}        %
\usepackage{braket}     %
\usepackage{makecell}
\usepackage{bbm}

\newcommand{\dd}{\text{d}}

\newcommand{\nn}{\nonumber}

\DeclareMathAlphabet{\mathdutchcal}{U}{dutchcal}{m}{n}
\SetMathAlphabet{\mathdutchcal}{bold}{U}{dutchcal}{b}{n}
\DeclareMathAlphabet{\mathdutchbcal}{U}{dutchcal}{b}{n}

\begin{document}

\title{Factorization and Resummation for the Nearside Energy-Energy Correlators}

\author{Yuxun Guo}
\email{yuxunguo@lbl.gov}
\affiliation{Physics Department, University of California, Berkeley, California 94720, USA}
\affiliation{Nuclear Science Division, Lawrence Berkeley National
Laboratory, Berkeley, CA 94720, USA}

\author{Feng Yuan}
\email{fyuan@lbl.gov}
\affiliation{Nuclear Science Division, Lawrence Berkeley National
Laboratory, Berkeley, CA 94720, USA}

\author{Wenbin Zhao}
\email{wenbinzhao@lbl.gov}
\affiliation{Physics Department, University of California, Berkeley, California 94720, USA}
\affiliation{Nuclear Science Division, Lawrence Berkeley National
Laboratory, Berkeley, CA 94720, USA}

\begin{abstract}
By utilizing the di-hadron fragmentation formalism, we extend the previous factorization of nearside energy-energy correlators (EEC) in the collinear limit and derive an all order resummation in the Fourier transform $b_T$-space. A perfect matching is obtained when we compare to the fixed-order results. We further demonstrate the resummation effects for the EEC in $e^+e^-$ annihilation and show that they will significantly improve the theoretical predictions at small angles.
\end{abstract}

\maketitle

\section{Introduction}

Energy-energy correlators (EECs) in $e^+e^-$ annihilation~\cite{Basham:1978bw,Basham:1977iq,Basham:1978zq,PLUTO:1985yzc,PLUTO:1979vfu,CELLO:1982rca,JADE:1984taa,Fernandez:1984db,Wood:1987uf,TASSO:1987mcs,AMY:1988yrv,TOPAZ:1989yod,ALEPH:1990vew,L3:1991qlf,L3:1992btq,DELPHI:1990sof,OPAL:1990reb,OPAL:1991uui,SLD:1994idb} and hadronic collisions have played an important role in studying the strong interaction physics at the frontier~\cite{Collins:1981uk,Ali:1982ub,Clay:1995sd,deFlorian:2004mp,DelDuca:2016csb,Tulipant:2017ybb,Kardos:2018kqj,Moult:2018jzp,Dixon:2018qgp,Dixon:2019uzg,Ebert:2020sfi,Schindler:2023cww}. Significant progress has been made in recent years ~\cite{Lee:2006nr,Berger:2003iw,Hofman:2008ar,Chen:2020vvp,Lee:2022ige,Craft:2022kdo,Komiske:2022enw,Andres:2022ovj,Andres:2023xwr,Andres:2023ymw,Yang:2023dwc,Andres:2024ksi,Barata:2023bhh,Barata:2023zqg,Bossi:2024qho,Lee:2023tkr,Lee:2023xzv,Lee:2024esz,Chen:2024nyc,Holguin:2023bjf,Holguin:2024tkz,Xiao:2024rol,Xing:2024yrb,Andres:2024hdd,Liu:2024lxy,Alipour-fard:2024szj,Kang:2024dja,Barata:2024wsu,Csaki:2024zig,Fu:2024pic,Apolinario:2025vtx,Barata:2025fzd,Chen:2025rjc,Moult:2025nhu,CMS:2024mlf,ALICE:2024dfl,Tamis:2023guc,CMS:2025ydi,ALICE:2025igw,ALICEpA}. The EEC not only provides a great opportunity to explore the quantum chromodynamics (QCD), but may also shed light on the hadronization effects in high-energy processes in $e^+e^-$ annihilation and hadronic collisions including proton-proton, proton-nucleus and nucleus-nucleus collisions at the LHC , RHIC and future EIC, see, a recent review in Ref.~\cite{Moult:2025nhu}.

Formally, the EEC is defined as
\begin{equation}
\frac{1}{\sigma_{\text{tot}}} \frac{\dd\Sigma}{\dd\cos\chi} = \frac{1}{\sigma_{\text{tot}}} \sum_{i,j} \int \dd\sigma^{ij} \, \frac{E_i E_j}{Q^2} \, \delta(\cos\chi - \hat{n}_i \cdot \hat{n}_j),
\end{equation}
where $\dd\sigma^{ij}$ denotes the semi-inclusive differential cross-section of two hadrons $h_i$ and $h_j$, with energies $E_i$ and $E_j$, and in the direction of the unit vectors $\hat{n}_i$ and $\hat{n}_j$, respectively. The total  energy is $Q = \sum_i E_i$, and $\chi$ is the angle between the two directions. For convenience, we define the variable $z \equiv (1 - \cos\chi)/2$. For the EEC in $e^+e^-$ annihilation, two special kinematics have been focused on in the studies: near-side with $z\to 0$ and away-side $z\to 1$.

In the back-to-back region with total transverse momentum $q_T$ of two particles much smaller than the total energy $Q$, the away side is dominated by soft and collinear gluon radiations and an appropriate transverse momentum dependent (TMD) factorization has been applied~\cite{Collins:1981uk,Collins:1981uw,Collins:1981va,Collins:1981zc} and all order resummation can be carried out~\cite{Collins:1984kg,deFlorian:2004mp}. The TMD resummation has been included in the phenomenology studies of the EEC measurements in $e^+e^-$ annihilation~\cite{deFlorian:2004mp,Tulipant:2017ybb,Kardos:2018kqj,Ebert:2020sfi,Kang:2024dja}. On the other hand, in the near-side region, the EEC is dominated by collinear gluon radiations and a resummation of large logarithms can be further included~\cite{Abbate:2010xh,Moult:2018jzp}. Additional contributions from power corrections also play important role in this region~\cite{Korchemsky:1999kt,Abbate:2010xh,Schindler:2023cww,Lee:2024esz,Chen:2024nyc}. However, in the end-point region where $z$ is very small, one has to go beyond the above formalism. This region is usually referred as the ``free hadrons" region as suggested from the experimental data~\cite{Komiske:2022enw}. Recently, a non-perturbative TMD fragmentation function has been proposed to describe the distribution in this region~\cite{Liu:2024lxy,Barata:2024wsu}.

In this paper, we investigate the di-hadron fragmentation function contributions~\cite{Konishi:1978yx,Konishi:1979cb,Vendramin:1981te,deFlorian:2003cg,Majumder:2004wh,Majumder:2004br,Jaffe:1997hf,Bianconi:1999cd,Bacchetta:2002ux,Bacchetta:2012ty,Zhou:2011ba,Cocuzza:2023vqs,Pitonyak:2023gjx} to the EEC observables. In particular, we will study the factorization and resummation for the nearside EEC utilizing this idea. In a recent paper, this has also been applied to study the scaling and scaling violation behavior depending on the hard momentum scale $Q$ by connecting them to the lightray operator product expansion (OPE)~\cite{Hao_Chen_future}.

Our formalism extends the previous factorization result for the EEC in the collinear limit~\cite{Dixon:2019uzg}. Based on this powerful argument of the collinear factorization, we focus on the small-angle EEC, where the distance between the two particles can be regarded as a transverse momentum difference in the transverse plane perpendicular to the jet axis. Therefore, a TMD description is a natural choice to perform the resummation and match to non-perturbative contributions~\cite{Liu:2024lxy,Barata:2024wsu}. Different from the arguments used in these earlier papers, here, we show explicitly how the factorization works, and an all order resummation is derived as a result of the factorization and the associated evolution equations.

The nearside factorization is different from the awayside factorization, although they both apply the TMD concept. In the awayside region, the TMD fragmentation function contains both collinear and soft gluon radiation contributions and a relevant Collins-Soper evolution is applied~\cite{Collins:1981uk,Collins:1981uw,Collins:1981va,Collins:1981zc}. On the other hand, in the nearside region, there is no soft gluon radiation, and the QCD evolution follows the exact DGLAP collinear evolution. In particular, there is no Sudakov-type of double logarithms at higher orders. This highlights a unique opportunity to study the non-perturbative QCD which is relevant to particle hadronization but different from traditional TMD physics discussed in the literature.

Meanwhile, the di-hadron fragmentation formalism has been extensively explored in hadron physics community to unveil novel nucleon property such as the nucleon's tensor charge~\cite{Jaffe:1997hf,Bianconi:1999cd,Bacchetta:2002ux,Bacchetta:2012ty,Zhou:2011ba,Cocuzza:2023vqs,Pitonyak:2023gjx}. In these studies, the associated fragmentation functions are usually parameterized and fitted to the experimental data. Our results in this paper will help constrain these functions and provide an additional input for phenomenological studies.

\section{Di-hadron Fragmentation in a Jet}

For any generic process in which two hadrons are produced, we have the following two contributions~\cite{deFlorian:2003cg},
\begin{equation}
    \frac{\dd\sigma(h_1h_2)}{\dd z_1 \dd z_2}=\sum_{ij}\hat\sigma_{ij}D_i^{h_1}(z_1)D_j^{h_2}(z_2) +\sum_i\hat \sigma_i I\!\!D_i^{h_1h_2}(z_1,z_2) \ ,\label{eq:dihadron}
\end{equation}
where $D_i^{h_1}(z_1)$ represents the fragmentation function describing the probability for parton $i$ to fragment into a hadron $h_1$ that carries a fraction $z_1$ of the parton's momentum, and similarly for $D_j^{h_2}(z_2)$. On the other hand, the di-hadron fragmentation function $I\!\!D_i^{h_1h_2}(z_1,z_2)$ describes two hadrons $h_1$ and $h_2$ with momentum fractions $z_1$ and $z_2$, respectively, coming from a single parton $i$. All these fragmentation functions and the partonic cross sections $\hat\sigma_{ij}$ and $\hat\sigma_i$ depend on the scales, which have been omitted for simplicity.
In the above equation, the first term gives the two individual fragmentation functions from two partons ($i$ and $j$) and the second gives the di-hadron fragmentation function from a single parton $i$. The latter contribution becomes important when the two hadrons are close to each other, where a collinear divergence from the first term needs to be absorbed into the renormalization of the di-hadron fragmentation function as well~\cite{deFlorian:2003cg}. In perturbative calculations, this divergence has to be properly taken care of so that the final results are free of divergence.

In addition, the momentum conservation leads to the following identity for the di-hadron fragmentation function~\cite{Konishi:1979cb,deFlorian:2003cg},
\begin{equation}
    \int \dd z_1 z_1\sum_{h_1}I\!\!D_i^{h_1h_2}(z_1,z_2)=(1-z_2)D_i^{h_2}(z_2) \ .\label{eq:sumrule}
\end{equation}
Applying the di-hadron fragmentation contribution to the EEC inside a high energy jet, we define the EEC jet function,
\begin{equation}
    \Gamma_i(\mu)=\sum_{h_1h_2}\int \dd z_1\dd z_2 z_1 z_2 I\!\!D_i^{h_1h_2}(z_1,z_2,\mu) \ ,
\end{equation}
which is directly connected to the EEC measurement in jet as we will show below. The EEC inside the jet is defined as~\cite{Komiske:2022enw},
\begin{eqnarray}
\label{eq:eecdef}
&&\frac{\dd
\langle {\rm EEC}(\theta)\rangle
}{\dd ^2{\theta}}   =   \frac{1}{N_{jet}}
\sum_{i\neq j\in {\rm J}}
\frac{E_iE_j}{E_J^2}  \delta^{(2)}(\vec{\theta}-(\vec{\theta}_i-\vec{\theta}_j)) \ ,\nn
\end{eqnarray}
where $N_{jet}$ represents the number of jets.
The sum is over all particles $(i,j)$
with relative angle $\theta$ and the jet energy $E_J=\sum_{i}E_i$.
Integrating over $\theta$, we define the integrated EEC,
\begin{eqnarray}
\langle {\rm EEC}\rangle (E_J)=
\frac{1}{N_{jet}}\sum_{jets\, {\rm J}}
\sum_{i\neq j\in {\rm J}}
\frac{E_iE_j}{E_J^2}  \ .\label{eq:eecintegral}
\end{eqnarray}
Focusing on the collinear kinematics of two particle production and applying the similar factorization argument of Ref.~\cite{Kaufmann:2015hma}, we can compute the above quantity in terms of the di-hadron fragmentation function,
\begin{eqnarray}
    \langle {\rm EEC}\rangle (E_J)&=& \sum_{i,h_1,h_2}\int \dd x x^2H_i(x,Q/\mu) \nn\\
    &&\times \int \dd z_1\dd z_2 z_1 z_2 I\!\! D_i^{h_1h_2}(z_1,z_2,\mu) \nn\\
    &=&\sum_i\int \dd x x^2H_i(x,Q/\mu) \Gamma_i(\mu) \label{eq:dihadron0}\ ,
\end{eqnarray}
where $H_i$ is the perturbative coefficient for parton $i$, carrying momentum fraction $x$ of the jet. The above factorization is similar to that of a semi-inclusive observable in jet~\cite{Kaufmann:2015hma,Chien:2015ctp,Gao:2024dbv,Neill:2016vbi,Kang:2017glf,Kang:2019ahe,XHLiu_future}. We take the leading order and collinear limit for the EEC in jet, while the first term of Eq.~(\ref{eq:dihadron}) will be included at higher orders, as we will show below through the QCD evolution. Here $Q$ represents the hard momentum scale $Q=E_{J}$.
At this order, the integrated EEC measures $\Gamma_q(\mu)$ and $\Gamma_g(\mu)$ for the quark and gluon jet, respectively. At higher order, there will be mixing between them, an important contribution to which comes from the evolution. The evolution of $\Gamma_i$ can be simplified by applying the momentum sum rule in Eq. (\ref{eq:sumrule}), %
\begin{equation}
    \Gamma_i(\mu)=1-\Gamma_i'(\mu) \ , \label{eq:gamma0}
\end{equation}
where $\Gamma_i'(\mu)$ is related to the third moments of the single hadron fragmentation functions
\begin{equation}
    \Gamma_i'(\mu)=\sum_h\int \dd z z^2 D_i^h(z,\mu)\ .
\end{equation}
With the above equations, it is straightforward to derive the evolution equations for $\Gamma_i$. Because of the mixing, it is appropriate to write down the evolution equation in the flavor-space matrix and we have, %
\begin{equation}
\frac{\dd}{\dd \ln{\mu^2}} \boldsymbol\Gamma=-\boldsymbol\Gamma\cdot\hat{{\boldsymbol{\gamma}}_T}(N) +\boldsymbol{1}\cdot\hat{{\boldsymbol{\gamma}}_T}(N)\ ,
\end{equation}
where $\boldsymbol\Gamma$ represents $(\Gamma_q,\Gamma_g)$ and $N=3$. It is interesting to find out that the second term is the inhomogeneous contribution which comes from the first term in Eq.~(\ref{eq:dihadron}). The above evolution can be verified by applying the scale evolution for the dihadron fragmentation functions~\cite{deFlorian:2003cg} in Eq.~(\ref{eq:dihadron0}). The anomalous dimensions $\hat{{\boldsymbol{\gamma}}_T}(N)$ are defined
\begin{equation}
    \hat{\boldsymbol{\gamma}}_T(N)\equiv- \int \dd x \ x^{N-1} \hat{\boldsymbol{P}}_T(x)\ ,\label{eq:gammant}
\end{equation}
with the time-like DGLAP splitting kernel $\hat{{\boldsymbol{P}}_T}$ as a flavor-space matrix,
\begin{equation}
    \hat{{\boldsymbol{P}}_T}=\begin{pmatrix}
    P_{qq} & P_{q g}\\
     P_{gq} & P_{gg}
    \end{pmatrix}\ .
\end{equation}
In practice, we can solve the evolution equation for $\Gamma'$ and make predictions for the integrated EEC, and we find that this gives a very good description of the integrated jet EEC of Eq. (\ref{eq:eecintegral}) at the LHC~\cite{Wenbin_Zhao_future}. %

\section{IR Behavior at Small Angle and the Factorization in Fourier Transform $b_T$-space}

Now, we turn to the angular distribution of the EEC. It has been realized that the small-angle EEC depends on non-perturbative physics. Here, we study its IR behavior by applying the EEC jet function constructed from the di-hadron fragmentation functions and derive a factorization formalism. This factorization extends the previous derivation in \cite{Dixon:2019uzg} in the collinear limit. A similar idea has been suggested in Ref.~\cite{Barata:2024wsu}. Our framework is a complete answer to the small angle resummation.

Let us start with the unintegrated EEC jet function $\Gamma_i(\mu,q_\perp)$, which can be defined through the di-hadron fragmentation functions, similar to that of Eq.~(\ref{eq:gamma0}).
To do that, we keep the transverse momenta for the two hadrons $k_{1\perp}$ and $k_{2\perp}$, respectively, and define $\vec{q}_\perp=\vec{k}_{1\perp}/z_1-\vec{k}_{2\perp}/z_2$. With this unintegrated EEC jet function, the angular distribution of EEC can be written as
\begin{equation}
    \frac{\dd \langle {\rm EEC}(\theta)\rangle (E_J)}{\dd ^2\theta}= E_J^2\sum_i\int \dd x x^4H_i(x,Q/\mu) \Gamma_i(\mu,q_\perp) \ ,\label{eq:dihadron1}
\end{equation}
where $q_\perp=xE_J\theta$.
In the transverse plane perpendicular to the jet direction, the angular distribution between the two particles comes from their transverse momentum distribution. Therefore, an analogous factorization in terms of transverse-momentum-dependent distributions is plausible. In the following, we show that is exactly the case.

\begin{figure}[htbp]
  \begin{center}
   \includegraphics[scale=0.37]{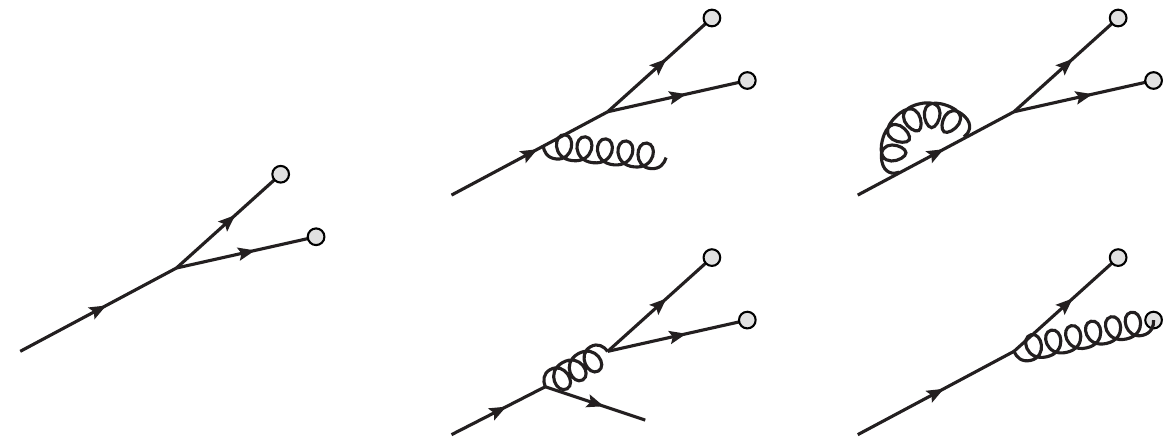}
\caption{Diagrammatic representations of the leading order and next-to-leading order contributions to the EEC calculations. The last diagram represents the inhomogeneous contribution, while the rest are homogeneous contributions because they do not change the transverse momentum difference between the two final state particles. }
  \label{fig:eec_spliting}
 \end{center}
  \vspace{-5.ex}
\end{figure}

In order to investigate the IR behavior for $\Gamma(\mu,q_\perp)$, we carry out a perturbative calculation, see, e.g., diagrams in Fig.~\ref{fig:eec_spliting}. At the leading order, we have a simple Delta function,
\begin{equation}
    \Gamma^{(0)}_q(\mu,q_\perp)=\Gamma_q(\mu)\delta^{(2)}(q_\perp) \ ,
\end{equation}
for the quark part, where $\Gamma_q(\mu)$ represents the integrated distribution. At one-loop order, similar to the above integrated EEC jet function, it receives homogeneous and inhomogeneous contributions. The homogeneous term is also proportional to a Delta function of $q_\perp$,
\begin{equation}
    \Gamma^{(1)}_q%
    |_{homo.}=\delta^{(2)}(q_\perp)%
    \left(-\frac{1}{\epsilon}+\ln\frac{\mu^2}{\bar\mu^2}\right)\Gamma_q(\bar\mu)(-\gamma_{qq}^{(3)})\ ,\label{eq:homo}
\end{equation}
where $\gamma_{qq}^{(3)}$ represents the $N=3$ anomalous dimension, see, Eq.~(\ref{eq:gammant}). Similarly, we should have gluon channel contribution of $\gamma_{gq}^{(3)}$.

The inhomogeneous term is the most important contribution, because it generates a nonzero $q_\perp$ (nonzero $\theta$ as well). At this order, it is straightforward to find out,
\begin{equation}
    \Gamma^{(1)}_q%
    |_{inhomo.}=\frac{\alpha_sC_F}{2\pi^2}\frac{1}{q_\perp^2}\frac{3}{2}\ ,\label{eq:inhomo}
\end{equation}
where we have taken into account both $qg$ and $gq$ configurations from the quark splitting. At this order, it is easy to check that,
\begin{equation}
    \gamma_{qq}^{(3)}+\gamma_{gq}^{(3)}=3  C_F \ .
\end{equation}
This leads to consistent evolution for the integrated EEC jet function after integrating over $q_\perp$.

Both equations of (\ref{eq:homo}) and (\ref{eq:inhomo}) contain IR divergences, and their total contributions lead to the IR behavior for the integrated EEC jet function $\Gamma_q(\mu)$. The resummation for finite $q_\perp$ can be built on this important feature. In previous studies~\cite{Konishi:1978yx,Konishi:1979cb,Vendramin:1981te,Dixon:2019uzg}, an integral over $q_\perp$ (or $z$) was taken and the associated resummation was carried out with a differential respect to $q_\perp$ after all order resummation. In the following, we take a different approach, following an analogous factorization argument as that for the TMD fragmentation functions.

The key element of our treatment is the Fourier transform to $b_T$-space,
\begin{equation}
    \widetilde{\Gamma}_q(\mu;b_T)=\int{\dd ^2q_\perp} e^{iq_\perp\cdot b_T}\Gamma(\mu;q_\perp) \ ,
\end{equation}
and factorization / resummation is performed in the $b_T$-space. The final result for the $q_\perp$ (and $\theta$) distribution is obtained by Fourier transforming back to $q_\perp$-space.

To illustrate the above point, let us start again a perturbative calculation in $b_T$ space, and the leading order result for $\widetilde{\Gamma}$ is simple,
\begin{equation}
    \widetilde{\Gamma}^{(0)}_q(\mu,b_T)=\Gamma_q(\mu) \ ,
\end{equation}
and one loop homogeneous term takes the same form,
\begin{equation}
    \widetilde{\Gamma}^{(1)}_q%
    |_{homo.}=%
    \left(-\frac{1}{\epsilon}+\ln\frac{\mu^2}{\bar\mu^2}\right)\left[-\Gamma_q(\bar\mu)\gamma_{qq}^{(3)}-\Gamma_g(\bar\mu)\gamma_{gq}^{(3)}\right]\ ,\label{eq:homobt}
\end{equation}
where we have included the contributions from both $q\to g$ and $q\to g$ channels. The inhomogeneous term comes again from the last diagram in Fig.~\ref{fig:eec_spliting}. By Fourier transforming Eq.~(\ref{eq:inhomo}) to $b_T$-space, we have
\begin{equation}
    \widetilde{\Gamma}^{(1)}_q%
    |_{inhomo.}=%
    \left(-\frac{1}{\epsilon}+\ln\frac{\mu_b^2}{\bar\mu^2}\right)\left(\gamma_{qq}^{(3)}+\gamma_{gq}^{(3)}\right)\ ,\label{eq:inhomobt}
\end{equation}
where $\mu_b=2e^{-\gamma_E}/b_T$ with $\gamma_E$ the Euler constant. Adding them together, we have
\begin{eqnarray}
   && \widetilde{\Gamma}^{(1)}_q(\mu,b_T)=\left(-\frac{1}{\epsilon}+\ln\frac{\mu_b^2}{\bar\mu^2}\right)\left[-\Gamma_q(\bar\mu)\gamma_{qq}^{(3)}-\Gamma_g(\bar\mu)\gamma_{gq}^{(3)}\right.\nn\\
    &&\left.+\gamma_{qq}^{(3)}+\gamma_{gq}^{(3)}\right]-\ln\frac{\mu^2}{\mu_b^2}\left[\Gamma_q(\mu_b)\gamma_{qq}^{(3)}+\Gamma_g(\mu_b)\gamma_{gq}^{(3)}\right]\ .
    \label{eq:oneloopbt}
\end{eqnarray}
In the above equation, the first term contains IR divergence which will be renormalized by the evolution of $\Gamma_q(\mu)$ at one-loop order. $b_T$ dependence only comes from the second term. %
The above demonstrates that we can factorize the EECs at small angles in terms of the EEC jet function which is constructed from the di-hadron fragmentation functions.

\section{All Order Resummation and Compare to the fixed-order Results}

After renormalization of the EEC jet functions, from the above one-loop calculations, we obtain the following result,
\begin{eqnarray}
   && \widetilde{\Gamma}^{(1)}_i(\mu,b_T)=\Gamma_i(\mu_b)-\ln\frac{\mu^2}{\mu_b^2}\gamma_{ji} \Gamma_j(\mu_b) \ ,
\end{eqnarray}
which applies to both quark and gluon cases and the EEC jet functions on the right-hand side are renormalized ones. An all order resummation can be derived by solving the scale evolution respect to the scale $\mu$. Following a similar method in~\cite{Dixon:2019uzg}, it can be solved as,
\begin{eqnarray}
\widetilde{\Gamma}_i(\mu,b_T)&=&\Gamma_j(\mu_b) \text{P}\exp\left[-\int_{\mu_b}^\mu\dd\ln\mu'^2 \gamma(\mu')\right]_{ji} \nn \\
&=&\Gamma_j(\mu_b)\text{P}\exp\left[\int_{\alpha_S(\mu_b)}^{\alpha_S(\mu)}\frac{\dd\alpha_S}{\alpha_S} \frac{\gamma(\alpha_S)}{\beta(\alpha_S)}\right]_{ji}\ ,
\end{eqnarray}
where $P$ implies a path-order matrix product in the integrating variable for the exponential term. The beta function $\beta\equiv-\dd\alpha_S(\mu)/\dd\ln\mu^2$ has been defined, and can be expanded as $\beta=\sum_n a_S^{n+1} \beta_n$. In this work, we define the perturbative expansion parameter $a_S\equiv \alpha_S/(4\pi)$.

It is not surprising that the above evolution kernel is identical to that derived previously~\cite{Dixon:2019uzg}, because the collinear divergence is the same. The difference here is that we perform the resummation in the Fourier transform $b_T$-space. It is interesting to note that the above result is similar to that suggested in~\cite{Barata:2024wsu}. In our derivation, the factorization and explicit calculation at one-loop order help identify the initial input for the evolution.

When Fourier transformed back to $q_\perp$-space, the above resummation formula will encounter the non-perturbative region. We follow the so-called $b_*$ prescription, $b_T\to b_*=b_T/\sqrt{1+b_T^2/b_{max}^2}$ where we choose $b_{max}=1.5\ \rm GeV^{-1}$. With that, we will also apply a non-perturbative contribution. The final resummation result for $\widetilde{\Gamma}$ reads as
\begin{equation}
    \widetilde{\Gamma}_{res.} (\mu,b_T)=\widetilde{\Gamma}_{perp}(\mu,b_*)e^{-g_{q,g}b_T^2} \ ,\label{eq:nonpert}
\end{equation}
where $g_q$ and $g_g$ are free parameters and will be fitted to experimental data.

The above resummation formula can be compared to the asymptotic behavior from the fixed-order calculations~\cite{Dixon:2018qgp,Dixon:2019uzg} and we find a perfect agreement, see, Appendix. This provides an important cross-check for our results and leads to a matching between the resummation results and the fixed-order calculations when applying to the experimental data.

Now, we apply our resummation formula to the unintegrated EEC in $e^+e^-$ and demonstrate its importance in the small-angle region.  With our resummation result, it can be written as %
\begin{eqnarray}
    \frac{\dd \Sigma}{\sigma_0\dd  z}&|_{z\ll 1}=& 4\pi Q^2\int\frac{\dd ^2b_T}{(2\pi)^2}e^{i\chi Q b_T} H_q\widetilde{\Gamma}_q^{res.}(Q,b_T) \ ,
\end{eqnarray}
at the leading logarithmic approximation (LLA).

\begin{figure}[htbp]
  \begin{center}
   \includegraphics[scale=0.556]{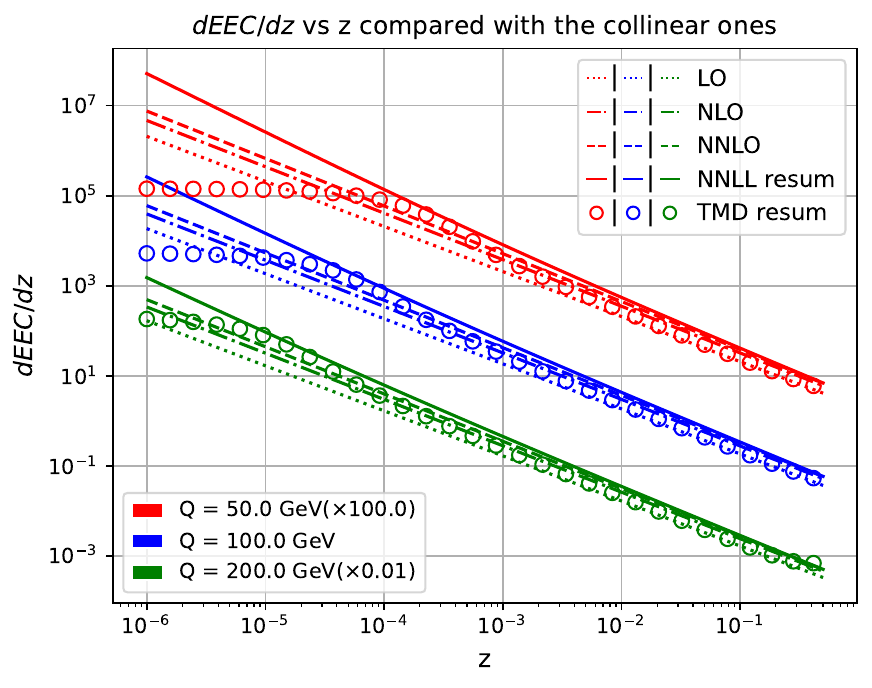}
\caption{EECs in $e^+e^-$ annihilation, comparing our resummation results with those from fixed-order calculations~\cite{Dixon:2018qgp} and the next-to-next-leading logarithmic (NNLL) resummation results from~\cite{Dixon:2019uzg} at small angles for different energies.
}
  \label{fig:eecee}
 \end{center}
  \vspace{-5.ex}
\end{figure}

In Fig.~\ref{fig:eecee}, we show the small-angle EECs as functions of $z$ for different energies, comparing the resummation results to the fixed-order calculations~\cite{Dixon:2018qgp,Dixon:2019uzg}. A number of important observations. First, at the moderate $z$, the resummation results match to the collinear calculations, and as expected, the agreement extends to smaller angles when we compare to the collinear resummation results from~\cite{Dixon:2019uzg}. Second, our TMD-type of resummation correctly predicts the turnover of the distributions at very small angles. This mainly comes from the non-perturbative contribution in our formalism, see, Eq.~(\ref{eq:nonpert}). In the numeric calculations, we assume the dominance of quark jets in $e^+e^-$ annihilation and take the leading-order hard part $H_q=1/2$. For the non-perturbative parameters, we take $b_{max}=1.5$ GeV$^{-1}$ and $g_q=g_g=0.35$ which is consistent with the parametrization used in Ref.~\cite{Liu:2024lxy}. We also utilize the single-hadron fragmentation functions \cite{Gao:2024nkz} at $\mu_0=2$ GeV and calculate $\Gamma'_q(\mu_0)=0.246$ and $\Gamma'_g(\mu_0)=0.176$ so that $\Gamma_i(\mu_0)=1-\Gamma'_i(\mu_0)=(0.754,0.824)$ set the initial condition of the evolution. Though we note that the agreement with the collinear calculations persists when varying these non-perturbative parameters, which affects mostly the plateau regions at very small angles.

\section{Conclusion}

In summary, by utilizing the di-hadron fragmentation function contributions to the EEC observables, we extend previous factorization and resummation formalism to the small-angle region. A TMD-type of resummation formula was derived following the factorization argument and the evolution equations for the EEC jet functions. Explicit results have been shown at the one-loop order. Asymptotic expansion has also been checked against the fixed-order calculations, and a perfect agreement was found at the leading logarithmic level.

Comparing our resummation formula to the fixed-order calculations for the EECs in $e^+e^-$ annihilation demonstrates the perfect matching between them. In particular, our resummation captures some important features of the collinear resummation results from Ref.~\cite{Dixon:2019uzg}. In addition, our resummation results predict a turnover in the angular distribution at very small angles, which is consistent with previous calculations using the TMD fragmentation functions~\cite{Liu:2024lxy,Barata:2024wsu}, and the ALEPH data \cite{Bossi:2024qeu}. We plan to carry out a detailed phenomenological comparison between our resummation formula and the experimental data in the nearside EEC from various collision systems. This provides an important link between the ``free hadron" phase and the collinear region in the EEC observables. Extension to higher-point energy-energy correlators shall follow as well.

The consistent treatment of the IR behavior derived in this paper should be applied to the phenomenology of hadron physics as well. This will strengthen the theoretical prediction power in the associated studies to extract the nucleon property from the di-hadron fragmentation processes~\cite{Jaffe:1997hf,Bianconi:1999cd,Bacchetta:2002ux,Bacchetta:2012ty,Zhou:2011ba,Cocuzza:2023vqs,Pitonyak:2023gjx}. For example, the Collins-type of contribution has been computed in the large invariant mass region for the di-hadron fragmentation function~\cite{Zhou:2011ba}. This should be part of the inhomogeneous contribution to the spin-dependent  di-hadron fragmentation function and a similar resummation should be applied there. We will come back to the above issues in a future publication.

\textbf{\textit{Acknowledgment.}}
We thank Daniel de Florian, Felix Ringer and Werner Vogelsang for helpful discussions during the CFNS-INT program on ``Precision QCD with the Electron-Ion Collider", May 12-June 20, 2025. This work is supported by the Office of Science of the U.S. Department of Energy under Contract No. DE-AC02-05CH11231, and by the U.S. Department of Energy, Office of Science, Office of Nuclear Physics, within the framework of the Quark Gluon Tomography (QGT) and Saturated Glue (SURGE) Topical Theory Collaborations. W.B.Z is also supported by the National Science Foundation under grant number ACI-2004571 within the framework of the XSCAPE project of the JETSCAPE collaboration.

{\it Note added:} When this paper is finalized, we notice a similar resummation paper on this topic~\cite{Lee:2025okn}. The main conclusions and the numeric results agree with each other.

\bibliographystyle{apsrev4-1}
\bibliography{refs.bib}

\newpage

\appendix

\section{Resummation of EEC and perturbative matching with collinear calculations}

As discussed in the main text, the inhomogeneous term in the evolution of the di-hadron fragmentation functions generates the $q_\perp$-dependence, and therefore is sensitive to the transverse scale $\mu_\perp$ upon transverse integration. This naturally motivates one to take an alternative treatment similar to that for the TMDs by performing a Fourier transform to the $b_T$ space:
\begin{equation}
    \widetilde{\Gamma}_i(\mu,b_T)=\int\dd ^2q_\perp e^{iq_\perp\cdot b_T}\Gamma_i(\mu,q_\perp) \ ,
\end{equation}
Then one has,
\begin{equation}
    \tilde{\Gamma}^{(1)}_i(\mu,b_T)=-\left(-\frac{1}{\epsilon}+\ln\frac{\mu^2}{\bar\mu^2}\right)\gamma_{ji} \Gamma_j(\bar{\mu}) +\ln\frac{\mu_b^2}{\bar\mu^2}\sum_{j}\gamma_{ji}\ ,
\end{equation}
and $\mu_b\equiv 2e^{-\gamma_E}/b_T$. One can eventually rearrange it into
\begin{eqnarray}
   && \widetilde{\Gamma}^{(1)}_i(\mu,b_T)=\Gamma_i(\mu_b)-\ln\frac{\mu^2}{\mu_b^2}\gamma_{ji} \Gamma_j(\mu_b) \ .
\end{eqnarray}
This can be solved in the form
\begin{eqnarray}
\widetilde{\Gamma}_i(\mu,b_T)=\Gamma_j(\mu_b) \text{P}\exp\left[-\int_{\mu_b}^\mu\dd\ln\mu'^2 \gamma(\mu')\right]_{ji}\ ,
\end{eqnarray}
where $P$ implies a path-order matrix product of the exponential term in $\mu'$, which can be expressed equivlanetly in terms of the strong coupling as:
\begin{eqnarray}
   \widetilde{\Gamma}_i(\mu,b_T)=\Gamma_j(\mu_b) \text{P}\exp\left[\int_{\alpha_S(\mu_b)}^{\alpha_S(\mu)}\frac{\dd\alpha_S}{\alpha_S} \frac{\gamma(\alpha_S)}{\beta(\alpha_S)}\right]_{ji}\ ,
\end{eqnarray}
where $\beta\equiv-\dd\alpha_S(\mu)/\dd\ln\mu^2$ which can be expanded as $\beta=\sum_n a_S^{n+1} \beta_n$ and the P here implies a path-order matrix product of the exponential term in $\alpha_S$.

Now we consider the leading log approximation when $\mu\gg\mu_b$, and expand the $\widetilde{\Gamma}_i(\mu,b_T)$ as a power series of $L$ defined as $L\equiv a_S\ln\left(\mu^2/\mu_b^2\right)$. We define the leading log coefficients $A^{(n)}_{ji}$ according to,
\begin{eqnarray}
\widetilde{\Gamma}_i(\mu,b_T)=\sum_{j}\Gamma_j(\mu_b)\sum_n L^n A^{(n)}_{ji}\ ,
\end{eqnarray}
which can then be solved in matrix form as
\begin{align}
\boldsymbol{A}^{(0)}&=\mathbbm{1}\ ,\quad  \boldsymbol{A}^{(1)}=-\boldsymbol{\gamma}\ ,\\
\boldsymbol{A}^{(2)}&=\frac{1}{2}\left(-\beta_0\boldsymbol{\gamma}+\boldsymbol{\gamma}^2\right)\ , \\
\boldsymbol{A}^{(3)}&=\frac{1}{6}\left(-2\beta_0^2\boldsymbol{\gamma}+3\beta_0\boldsymbol{\gamma}^2-\boldsymbol{\gamma}^3\right)\ ,
\end{align}
where $\mathbbm{1}$ stands for the identity matrix and all the powers of $\gamma$ are matrix multiplications.

To compare it with the collinear calculation, we take the input $\boldsymbol{\Gamma}=(1,1)$ and also put in the hard part $\boldsymbol{H}=(1/2,0)$ for the EEC in $e^+e^-$ annihilation \footnote{The leading order hard coefficients are $\boldsymbol{H}(x)=(1/2\delta(1-x),0)$ and the integral $\int\dd x\  x^4 \boldsymbol{H}(x)$ gives $\boldsymbol{H}=(1/2,0)$}. Then we have the following leading log coefficients defined as $\bar{A}^{(n)}=\Gamma\cdot\boldsymbol{A}^{(n)}\cdot H  $ :
\begin{widetext}
\begin{align}
    \bar A^{(1)}&=-\frac{3}{2}C_F=-2\ ,\\
    \bar A^{(2)}&=\frac{C_F(-428C_A+375C_F+53n_f)}{120}=-\frac{173}{30}\ ,\\
    \bar A^{(3)}&= \frac{C_F \left(96708 C_A^2 - 2 C_A \left(68000 C_F + 16259 n_f \right) + 5\left(9375 C_F^2 + 4619 C_F\, n_f + 368 n_f^2 \right)\right)}{10800}=-\frac{20317}{1350}\ .
\end{align}
\end{widetext}
These results match perfectly with the leading log coefficients from collinear calculations~\cite{Dixon:2019uzg}, noting that a map between in the form of
\begin{equation}
    \frac{1}{z}\ln^n(z)\leftrightarrow \frac{(-1)^n}{n+1}\ln^{n+1}\left(\frac{\mu^2}{\mu_b^2}\right)
\end{equation}
noting that $z\equiv(1-\cos\chi)/2\propto q_T^2$. Thus a factor of $-1,2,-3,\cdots$ should be multiplied to the above coefficients when comparing with the collinear convention.

\end{document}